\documentclass[runningheads]{llncs}
\usepackage{epsfig}
\usepackage{graphicx}
\usepackage{amsmath}
\usepackage{amssymb}
\usepackage{subfigure}
\usepackage{multirow}
\usepackage{array}
\usepackage{color}
\usepackage{tabu}
\usepackage{longtable}

\usepackage{acro} 

\usepackage[breaklinks=true,colorlinks,bookmarks=false]{hyperref}
\newcommand{\etal}{\textit{et al}. }
\newcommand{\ie}{\textit{i}.\textit{e}., }
\newcommand{\eg}{\textit{e}.\textit{g}., }

\DeclareAcronym{hospital}{short=Anonymous, long={\it anonymous} hospital}  

\begin{document}
\title{Weakly-Supervised Universal Lesion Segmentation with Regional Level Set Loss}

\titlerunning{Weakly-Supervised ULS with Regional Level Set Loss}

%

\author{Youbao Tang\inst{1} \and
Jinzheng Cai\inst{1} \and
Ke Yan\inst{1} \and
Lingyun Huang\inst{2} \and
Guotong Xie\inst{2} \and
Jing Xiao\inst{2} \and
Jingjing Lu\inst{3} \and
Gigin Lin\inst{4} \and
Le Lu\inst{1}}

\institute{PAII Inc., Bathesda, MD, USA \\ \email{tybxiaobao@gmail.com; tiger.lelu@gmail.com} \and Ping An Technology, Shenzhen, PRC \and Beijing United Family Hospital, Beijing, PRC \and Chang Gung Memorial Hospital, Linkou, Taiwan, ROC}

\maketitle              

\begin{abstract}
Accurately segmenting a variety of clinically significant lesions from whole body computed tomography (CT) scans is a critical task on precision oncology imaging, denoted as universal lesion segmentation (ULS). Manual annotation is the current clinical practice, being highly time-consuming and inconsistent on tumor's longitudinal assessment. Effectively training an automatic segmentation model is desirable but relies heavily on a large number of pixel-wise labelled data. Existing weakly-supervised segmentation approaches often struggle with regions nearby the lesion boundaries. In this paper, we present a novel weakly-supervised universal lesion segmentation method by building an attention enhanced model based on the High-Resolution Network (HRNet), named AHRNet, and propose a regional level set (RLS) loss for optimizing lesion boundary delineation. AHRNet provides advanced high-resolution deep image features by involving a decoder, dual-attention and scale attention mechanisms, which are crucial to performing accurate lesion segmentation. RLS can optimize the model reliably and effectively in a weakly-supervised fashion, forcing the segmentation close to lesion boundary. Extensive experimental results demonstrate that our method achieves the best performance on the publicly large-scale DeepLesion dataset and a hold-out test set.

\keywords{Universal Lesion Segmentation \and Weakly-supervised Learning \and Regional Level Set Loss \and Computed Tomography.}
\end{abstract}
\section{Introduction}

Basing on global cancer statistics, $19.3$ million new cancer cases and almost $10.0$ million cancer deaths occurred in 2020 \cite{sung2021global}. Cancer is one of the critical leading causes of death and a notorious barrier to increasing life expectancy in every country of the world. To assess cancer progress and treatment responses, tumor size measurement in medical imaging and its follow-ups is one of the most widely accepted protocols for cancer surveillance \cite{eisenhauer2009new}. In current clinical practice, most of these measurements are performed by doctors or radiology technicians \cite{beaumont2019radiology}. It is time-consuming and often suffers from large inter-observer variations, especially with the growing cancer incidence. Automatic or semi-automatic lesion size measurement approaches are in need to 
alleviate doctors from this tedious clinical load, and more importantly, to significantly improve assessment consistency \cite{cai2018accurate,tang2020one}. In this work, we develop a new universal lesion segmentation (ULS) method to measure tumor sizes accurately on selected CT cross sectional images, as defined by RECIST guideline \cite{eisenhauer2009new}. 

Many efforts have been developed for automating lesion size measurement. Specifically, deep convolutional neural networks are successfully applied to segment tumors in brain \cite{havaei2017brain}, lung \cite{wang2017central,jin2018ct,zhou2019progressively}, pancreas \cite{zhu2019multi,zhang2020robust}, liver \cite{christ2017automatic,li2018h,chlebus2018automatic,raju2020co,tang2020e2net,ZHANG2021102005}, enlarged lymph node \cite{nogues2016automatic,zhu2020lymph}, \textit{etc}. Most of these approaches are specifically designed for a certain lesion type, however, an effective and efficient lesion size measurement tool should be able to handle a variety of lesions in practice \cite{cai2018accurate,tang2018ct,agarwal2020weakly,tang2020one}. Our ULS approach is proposed via leveraging a sophisticated network architecture and an effective weakly-supervised learning strategy. On one hand, more sophisticated network backbones allow ULS to have larger model capacities to cope with lesions with various appearances, locations and sizes. On the other hand, weakly-supervised learning strategy may drastically simplify the annotation complexity that permits large amounts of bookmarked cancer images (already stored in PACS system) to be used for model initialization. In weakly-supervised learning, we propose a new regional level set (RLS) loss as the key component to refine segment regions near lesion boundaries so as to improve the quality of segmentation supervisory signal.    

We follow the literature to formulate lesion size measurement as a two dimensional region segmentation problem, which performs dense pixel-wise classification on RECIST-defined CT axial slices. Such region segmentation based tumor size, area or volume assessment should perform more accurately in measuring solid tumor's response than lesion diameters \cite{eisenhauer2009new,tang2018semi,tang2020one}. The main reason why diameter is adopted in \cite{eisenhauer2009new} was due to the easier reproducibility by human raters. To precisely delineate the tumor boundary, we propose three main contributions in ULS: 1) an effective network architecture (AHRNet) based on HRNet \cite{wang2020deep} that renders rich high-resolution representations with strong position sensitivity, by being augmented with a decoder (DE) and a novel attention mechanism combining both dual attention (DA) \cite{fu2019dual} and scale attention (SA). 2) the RLS loss as a reformulated deep learning based level set loss \cite{chan2001active} with specific modifications for lesion segmentation. 3) AHRNet and the RLS loss integrated within a simple yet effective weakly-supervised training strategy so that our AHRNet model can be trained on large-scale PACS stored lesion databases, such as DeepLesion \cite{yan2018deeplesion}. 

Our contributions result in a new state-of-the-art segmentation accuracy that outperforms nnUNet \cite{isensee2018nnu} by 2.6\% in averaged Dice score and boosts the former best result \cite{tang2020one} from 91.2\% to 92.6\% in Dice on the DeepLesion test set. Our AHRNet model is also more robust by improving the worst case of 78.9\% Dice from 43.6\% in \cite{isensee2018nnu}. It is trained with a large-scale database and generalizes well on a hold-out test set, outpacing nnUNet by 2.7\% and achieving 88.2\% Dice score. Over 92\% of the testing lesions are segmented with  $>$85\% Dice scores, demonstrating that our AHRNet model is a reliable tool for lesion size measurement. The network components, \ie DE, DA, SA, and RLS, could work seamlessly with different network backbones including nnUNet, HRNet, and our AHRNet. We validate the effectiveness of each sub-module via ablation studies.  

\section{Methodology} \label{ahrnet}
\begin{figure*}[t!]
  \centering
  \includegraphics[width=\linewidth]{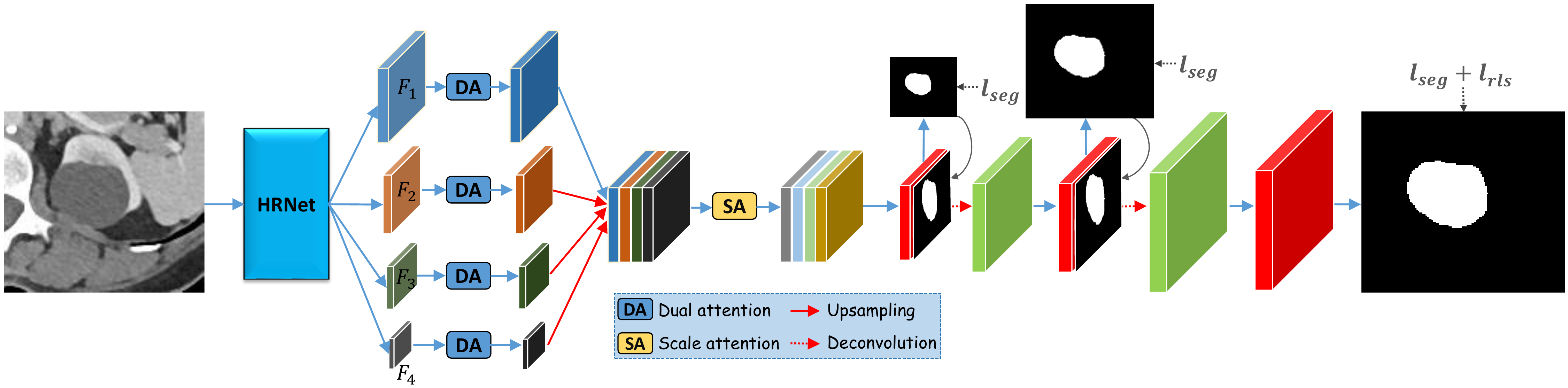}
  \caption{Illustrated framework of our proposed AHRNet, where $\ell_{seg}$ is the segmentation loss defined in Sec. \ref{sec:model-optimization} that consists of a binary cross entropy loss and an IoU loss, $\ell_{rls}$ is the regional level set loss described in Sec. \ref{sec:rls}.}
  \label{fig:framework} \vspace{-4mm}
\end{figure*}

This work aims to produce reliable and accurate lesion masks on the given lesion sub-images. Following previous work \cite{cai2018accurate,tang2020one}, we assume that the lesion sub-images have been obtained in the form of bounding boxes that could be either automatically generated by lesion detection algorithms or semi-automatically drawn by radiologists.
Fig. \ref{fig:framework} illustrates the overall AHRNet framework.

\subsection{AHRNet Architecture} \label{sec:ahrnet}
HRNet has been demonstrated of achieving state-of-the-art performance in a wide range of computer vision applications \cite{he2017mask}, including semantic segmentation, object detection, and human pose estimation, suggesting that HRNet may be a strong versatile CNN backbone. It can connect high and low resolution convolutions in parallel, maintain high resolution through the whole process, and fuse multi-resolution representations repeatedly, rendering rich hierarchical, high-resolution representations with strong position sensitivity. These characteristics of HRNet are crucial for pixel-wise dense prediction tasks. We choose HRNet as the backbone to extract rich multi-scale features for lesion segmentation here. As shown in Fig. \ref{fig:framework}, given a CT image $I \in \mathbb{R}^{H \times W}$, HRNet produces stacked multi-scale image features 
$F=\{F_k \in \mathbb{R}^{2^{k+4} \times \frac{H}{2^{k+1}} \times \frac{W}{2^{k+1}}}~|~k \in \{1,2,3,4\}\}$.

A straightforward means of lesion segmentation is to upsample $F_i$ to have the same resolution (\eg $\frac{1}{4}$ of the input image), concatenate them, and follow a convolutional layer with a 1$\times$1 kernel to get the prediction, which serves as our baseline. The resolution of deep image features is important for accurate lesion segmentation, especially for small lesions. Thus to get more accurate predictions, we set up a small decoder (DE) to obtain higher resolution features. From Fig. \ref{fig:framework}, it contains two deconvolutional layers with 32 4$\times$4 kernels and a stride of 2 and three convolutional layers with 32 3$\times$3 kernels, where the dimensions of feature maps are $\frac{1}{4}$, $\frac{1}{2}$, and $1$ of the input image, respectively. Another three convolutional layers with a 1$\times$1 kernel are added to get the corresponding predictions. Each deconvolutional layer takes as input the features and the prediction.

\begin{figure}[t!]
  \centering
  \includegraphics[width=0.7\linewidth]{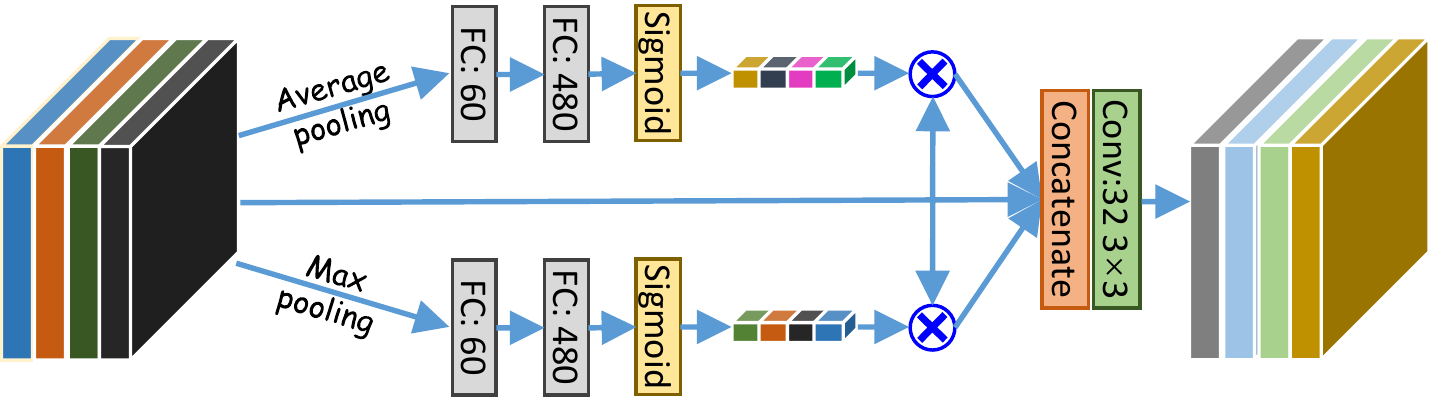}
  \caption{Overview of the scale attention module. It has two branches whose structures are similar to SE block \cite{hu2020squeeze}.}
  \label{fig:sa}
\end{figure}

As described above, we do not model the long-range dependencies of features in $F_i$ for lesion segmentation. However, long-range contextual information can be crucial in obtain precise predictions. Fu \etal \cite{fu2019dual} present a dual attention (DA) module that can capture the long-range contextual information over local feature representations in spatial and channel dimensions respectively. In this work, we model the global contextual information in $F_i$ by employing a DA  module \cite{fu2019dual} to adaptively aggregate their rich long-range contextual dependencies in both spatial and channel dimensions, and enhancing feature representations to improve the performance of lesion segmentation. Since the studied lesion sizes are very diverse, to better address lesion segmentation under different scales, we introduce a scale attention (SA) module to effectively combine the multi-scale features by treating them input-specifically with learnable scale attention vectors. It contains two branches that are built upon SE block \cite{hu2020squeeze}, as in Fig. \ref{fig:sa}.

\subsection{Regional Level Set Loss} \label{sec:rls}
A classic level set method is proposed for image segmentation \cite{chan2001active}, treating segmentation as an energy minimization problem. The energy function is defined as:
\begin{equation} \label{eq:ls}
    \begin{aligned}
        E\left(c_1, c_2, \phi\right) 
        &= \mu \cdot \operatorname{Length}(\phi) + \nu \cdot \operatorname{Area}(\phi) + 
        \lambda_1 \sum_{i \in I} |i - c_1|^2 H(\phi(i)) \\
        &+ \lambda_2 \sum_{i \in I} |i - c_2|^2 (1 - H(\phi(i))), 
    \end{aligned}
\end{equation}
where $\mu, \nu, \lambda_{1}$ and $\lambda_{2}$ are the predefined non-negative hyper-parameters, $i$ is the intensity of its corresponding image location, $\phi(\cdot)$ is the level set function, $\operatorname{Length}(\phi)$ and $\operatorname{Area}(\phi)$ are the regularization terms with respect to the length and the inside area of the contour, $c_1$ and $c_2$ represent the mean pixel intensity of inside and outside areas of the contour, and $H$ is the Heaviside function: $H(\phi(i))=1$, if $\phi(i) \geq 0$; $H(\phi(i))=0$ otherwise.

Recently, researchers have studied to integrate this energy function into deep learning frameworks for semantic segmentation \cite{kim2019cnn} and medical image segmentation \cite{chen2019learning,zhang2020deep}. Approaches \cite{kim2019cnn,chen2019learning} replace the original image $I$ in Eq. \ref{eq:ls} with a binary image that is reconstructed from the ground truth mask of each object. Method \cite{zhang2020deep} computes a cross-entropy loss between the outputs of Eq. \ref{eq:ls} when setting $\phi$ as the prediction and ground truth. Neither formulation applies for our task due to the lack of ground truth masks of lesions for training. To tackle this issue, based on Eq. \ref{eq:ls}, we introduce a regional level set (RLS) loss defined by
\begin{equation} \label{eq:lrs}
    \ell_{rls} = \frac{1}{|I'|}\sum_{i \in I'} [ \lambda_1 \cdot p(i) \cdot |i - c_1|^2  + \lambda_2 \cdot (1-p(i)) \cdot |i - c_2|^2],    
\end{equation}
where $p(i)$ is the predicted probability map of pixel $i$, $I'$ is the constrained region of the input image $I$ and $|I'|$ is the number of pixels in $I'$. We experimentally set $\lambda_1$$=$1, $\lambda_2$$=$3. Here, terms of $\operatorname{Length}(\phi)$ and $\operatorname{Area}(\phi)$ in Eq. \ref{eq:ls} have been removed because they are sensitive to object sizes (which vary greatly in our task). 

\begin{figure}[t!]
  \centering
  \includegraphics[width=0.9\linewidth]{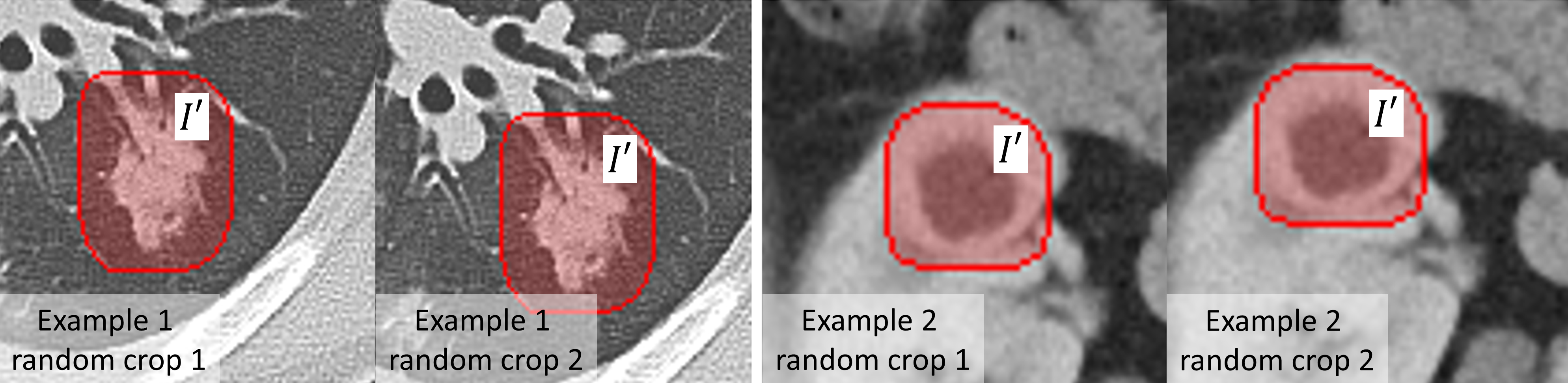}
  \caption{Two examples of using lesion-adaptive regions $I'$ for $\ell_{rls}$ computation defined in Eq. \ref{eq:lrs}.}
  \label{fig:rls}
\end{figure}

During training, we first obtain a lesion pseudo mask $g$ that is an ellipse for the given lesion image, fitted from four endpoints of its RECIST annotation, then construct the constrained region $I'$ by dilating $g$ to four times its size so that $I'$ is lesion-adaptive. Fig. \ref{fig:rls} shows two examples, where the region inside the red curve is the constrained region $I'$. As we can see, for the same lesion, the size of $I'$ remains stable under different data augmentations, \eg randomly cropping and rotating.

\subsection{Model Optimization} \label{sec:model-optimization}
As in Fig. \ref{fig:framework}, AHRNet takes as input a CT image and outputs three probability maps (denoted as $p_1$, $p_2$, and $p_3$). Besides the regional level set loss $\ell_{rls}$, we also use a segmentation loss ($\ell_{seg}$) to compute the errors between the predicted probability maps and the pseudo masks (denoted as $g_1$, $g_2$, and $g_3$) for optimization. $\ell_{seg}$ is the summation of a binary cross entropy loss ($\ell_{bce}$) and an IoU loss ($\ell_{iou}$), 
\ie $\ell_{seg}=\sum_{k=1}^{3}{[\ell_{bce}(p_k,g_k)+\ell_{iou}(p_k,g_k)]}$, which are defined as
\begin{equation}
    \begin{aligned}
        \ell_{bce}(p, g) 
        &= - \frac{1}{|I|} \sum_{i \in I} [g(i)\log(p(i)) + (1-g(i))\log(1-p(i))], \\
        \ell_{iou}(p, g) 
        &= 1 - \left(\sum_{i \in I} g(i)p(i)\right) / \left(\sum_{i \in I} g(i) + p(i) - g(i)p(i)\right), 
    \end{aligned}
\end{equation}
where we omit the subscript $k$ of $p$ and $g$ for simplicity. Although as a pixel-wise loss, $\ell_{bce}$ does not consider the global structure of lesion, $\ell_{iou}$ can optimize the global structure of the segmented lesion rather than focusing on a single pixel.

In order to make $\ell_{rls}$ to provide effective gradients for back propagation, we do not add $\ell_{rls}$ for training until the model converges using only $\ell_{seg}$. That means the model can produce a good-quality prediction at its early training stage with $\ell_{seg}$, which could be considered as a good initialization for $\ell_{rls}$. We then add $\ell_{rls}$, which is reduced by a factor of 0.1, at the later training stage so that it can provide useful gradients for optimization, making the prediction closer to the lesion boundary. 

The supervision for training is the constructed pseudo mask $g$ and its quality directly affects the final lesion segmentation performance. However, our straightforward ellipse estimation is not guaranteed to always generating lesion masks with high fidelity. Therefore, based on the prediction $p$ from the trained model and the fitted ellipse $e$, we further construct an updated pseudo mask $g'$ by setting $p$$\cap$$e$ as the foreground, $p$$\cup$$e$$-$$p$$\cap$$e$ as the ignored region, and the rest as the background. With the updated pseudo masks, we can retrain the model using the same way described above. The training has converged after three rounds.

\section{Experiments}

\textbf{Datasets and Evaluation Metrics.}
NIH DeepLesion dataset \cite{yan2018deeplesion} has $32,735$ CT lesion images from $4,459$ patients, where a variety of lesions over the whole body parts are included, such as lung nodules, liver lesions, enlarged lymph nodes, etc. Each lesion has only a RECIST annotation that serves as weak supervision for model optimization. Following \cite{cai2018accurate}, $1,000$ lesion images from 500 patients are manually segmented as a test set for quantitative evaluation. The rest patient data are used for training. 
Besides, we collect a hold-out test set from our collaborated \acl{hospital} for external validation. It contains 470 lesions from 170 patients with pixel-wise manual masks, which also covers various lesion types over the whole body.
The precision, recall, and Dice coefficient are used for performance evaluation.

\noindent \textbf{Implementation Details.}
AHRNet is implemented in PyTorch \cite{nips/PaszkeGMLBCKLGA19} and its backbone is initialized with \mbox{ImageNet} \cite{DBLP:conf/cvpr/DengDSLL009} pre-trained weights \cite{wang2020deep}. It is trained using Adam optimizer \cite{kingma2014adam} with an initial learning rate of 0.001 for 80 epochs reduced by 0.1 at epoch 40 and 60. The data augmentation operations include randomly scaling, cropping, rotating, brightness and contrast adjusting, and Gaussian blurring. After data augmentation, the long sides of all training images are randomly resized into a range of [128, 256]. For testing, an image is taken as input directly if its long side is in the range, otherwise, it is resized into the closest bound first. Although this work takes as input a lesion-of-interest region, it can cooperate with lesion detection and tracking techniques \cite{tang2019uldor,yan2019mulan,yan2020learning,cai2020deep,yan2020self} to perform automatic lesion segmentation on the entire CT images.

\begin{table}[t!]
	\begin{center}
		\caption{Lesion segmentation results of different methods. The mean and standard deviation of pixel-wise recall, precision and Dice score are reported (\%).}
		\label{tab:result}
		\scriptsize
		\begin{tabular}{|@{}*{1}{m{2.4cm}<{\centering}@{}}|@{}*{1}{m{1.6cm}<{\centering}@{}|@{}}*{1}{m{1.6cm}<{\centering}@{}|@{}}*{1}{m{1.6cm}<{\centering}@{}|@{}}*{1}{m{1.6cm}<{\centering}@{}|@{}}*{1}{m{1.6cm}<{\centering}@{}|@{}}*{1}{m{1.6cm}<{\centering}@{}|@{}}}
		\hline
		& \multicolumn{3}{c|}{DeepLesion test set} & \multicolumn{3}{c|}{Hold-out test set} \\ \cline{2-7}
        \multirow{-2}{*}{Method} & Precision & Recall & Dice & Precision & Recall & Dice \\ \hline
		\hline
		Cai \etal \cite{cai2018accurate}  &  89.3$\pm$11.1  &  93.3$\pm$9.5 & 90.6$\pm$8.9  & -  & - & -  \\ \hline
		Tang \etal \cite{tang2020one}  &  88.3$\pm$5.7  &  \textbf{94.7$\pm$7.4} & 91.2$\pm$3.9  & -  & - & -  \\ \hline
		nnUNet \cite{isensee2018nnu}  &  95.5$\pm$5.3  &  85.8$\pm$8.8 & 90.0$\pm$4.9  & 88.2$\pm$12.3 & 85.5$\pm$13.0 & 85.5$\pm$8.7   \\ 
		nnUNet+RLS  &  96.8$\pm$4.7  &  87.1$\pm$8.6 & 91.4$\pm$5.7  & 89.8$\pm$10.9 & 85.8$\pm$10.3 & 86.8$\pm$6.9   \\ \hline
		HRNet \cite{wang2020deep}  &  \textbf{97.5$\pm$3.2}  &  84.9$\pm$8.6 & 90.5$\pm$5.3  & 86.0$\pm$13.9 & 88.7$\pm$11.7 & 86.0$\pm$9.4  \\ 
		HRNet+RLS  &  95.0$\pm$5.8  &  89.7$\pm$9.4 & 91.8$\pm$6.2  & 86.9$\pm$12.1 & \textbf{90.3$\pm$10.4} & 87.6$\pm$8.1   \\ \hline
		AHRNet  &  97.0$\pm$3.7  &  87.0$\pm$8.3 & 91.5$\pm$5.1  & 88.5$\pm$11.3 & 87.7$\pm$11.7 & 86.8$\pm$6.4   \\ 
		AHRNet+RLS  &  95.8$\pm$4.5  &  90.2$\pm$7.4 & \textbf{92.6$\pm$4.3}  & \textbf{89.8$\pm$10.0} & 88.3$\pm$9.6 & \textbf{88.2$\pm$6.0}   \\ \hline
	    \end{tabular}
	\end{center}
	\vspace{-5mm}
\end{table}

\begin{figure}[t!]
    \centering
	\begin{minipage}[b]{0.49\linewidth}
		\centering
		\includegraphics[width=0.98\linewidth]{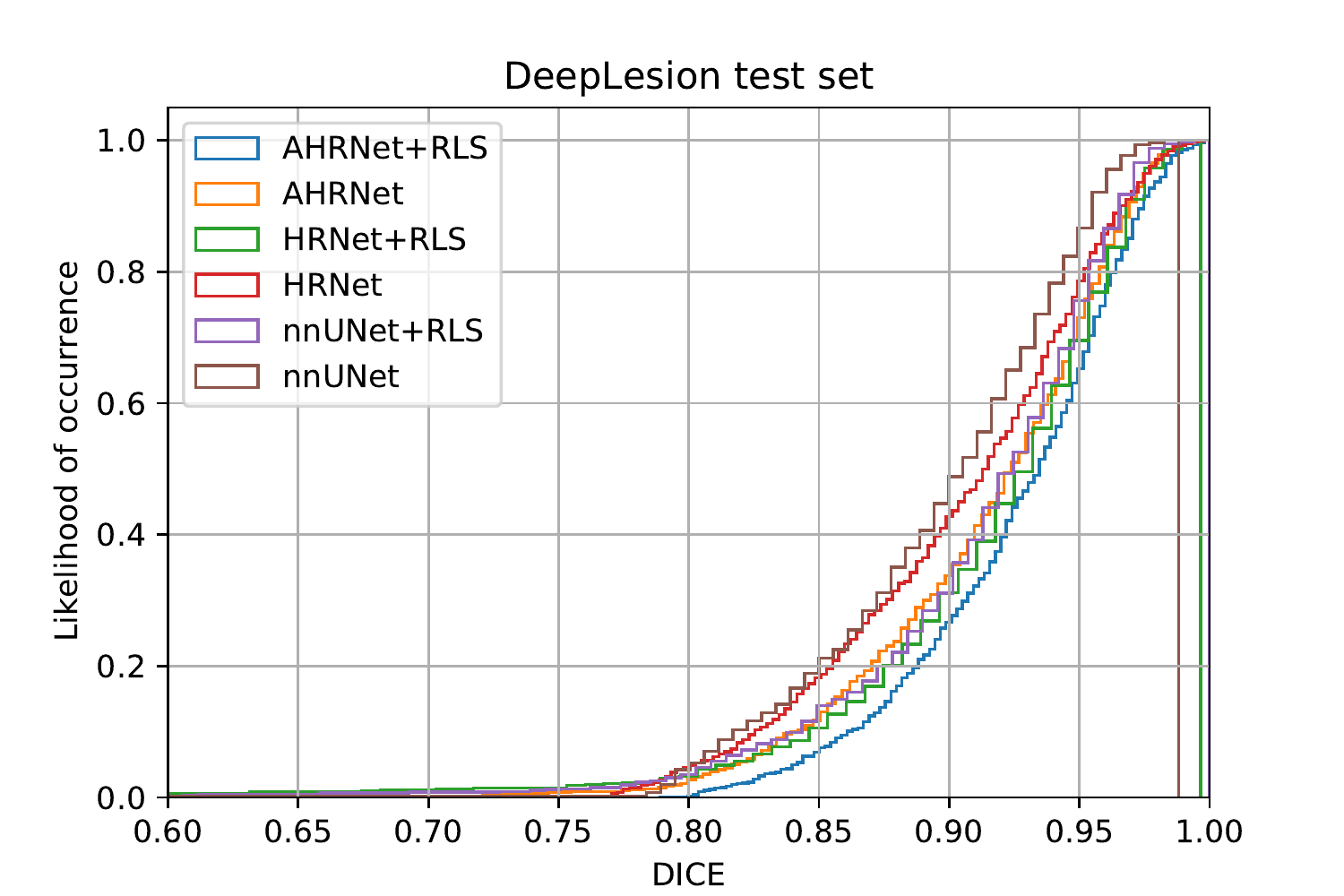} \\
	\end{minipage}
	\begin{minipage}[b]{0.49\linewidth}
		\centering
		\includegraphics[width=0.97\linewidth]{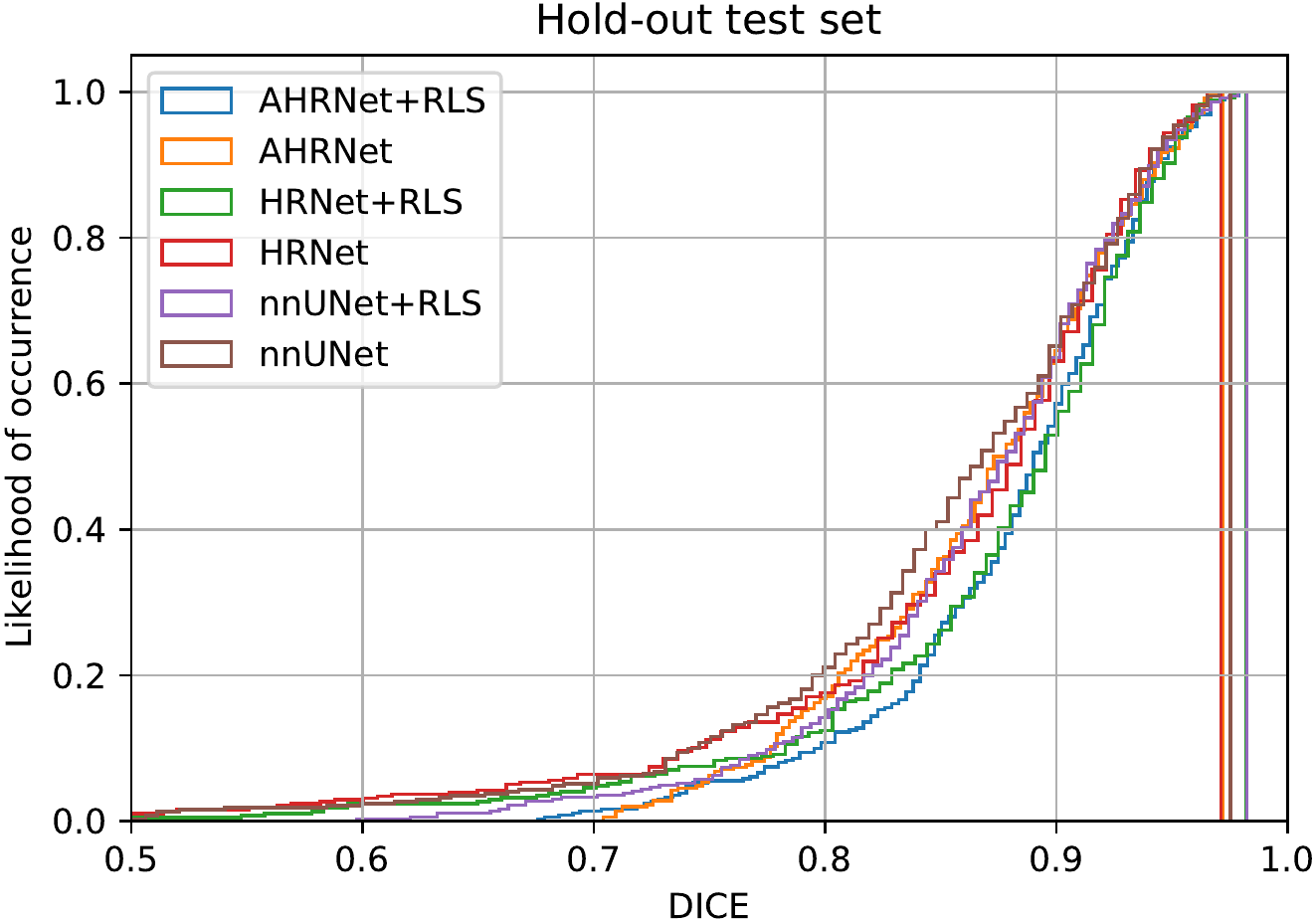} \\
	\end{minipage}
	\caption{Dice cumulative histograms of different methods on both test sets. The $x$ axes are truncated for better visualization when drawing the figures.}
	\label{fig:dice}
	\vspace{-5mm}
\end{figure}

\noindent \textbf{Quantitative Results.}
nnUNet \cite{isensee2018nnu} is a robust and self-adapting framework on the basis of vanilla UNets \cite{ronneberger2015u}. It has been widely used and overwhelmingly successful in many medical image segmentation tasks, suggesting itself as a strong baseline for comparisons. For empirical comparisons, besides the existing work \cite{cai2018accurate,tang2020one}, we train three segmentation models, \ie nnUNet, HRNet, and the proposed AHRNet, with or without our proposed RLS loss. Table \ref{tab:result} reports the quantitative results of different methods/variations on two test sets. As can be seen, 1) our method ``AHRNet+RLS'' achieves the highest Dice of 92.6\% surpassing the best previous work \cite{tang2020one} by 1.4\%, which demonstrates its effectiveness for weakly-supervised lesion segmentation. 2) When RLS is not used, the proposed AHRNet still has the best Dice score, indicating that the designed components are effective to enhance the feature representations for lesion segmentation. 3) For all three models, \ie nnUNet, HRNet, and AHRNet, the performance is consistently and remarkedly improved  when using RLS; the Dice gains are 1.4\%, 1.3\%, and 1.1\%, respectively. This shows that RLS is capable of making the segmentation outcome promisingly close to lesion boundaries and can be effectively optimized via a weakly-supervised fashion. Fig. \ref{fig:dice} shows the Dice cumulative histograms of different methods. Our method is observed with about 99\% or 90\% lesions having Dice $\geq$ 0.8 on the DeepLesion or hold-out test sets, respectively. Fig. \ref{fig:dice} evidently validates the overall improvements by our methods.

\begin{figure}[t!]
    \centering
	\begin{minipage}[b]{\linewidth}
		\centering
		\includegraphics[width=0.99\linewidth]{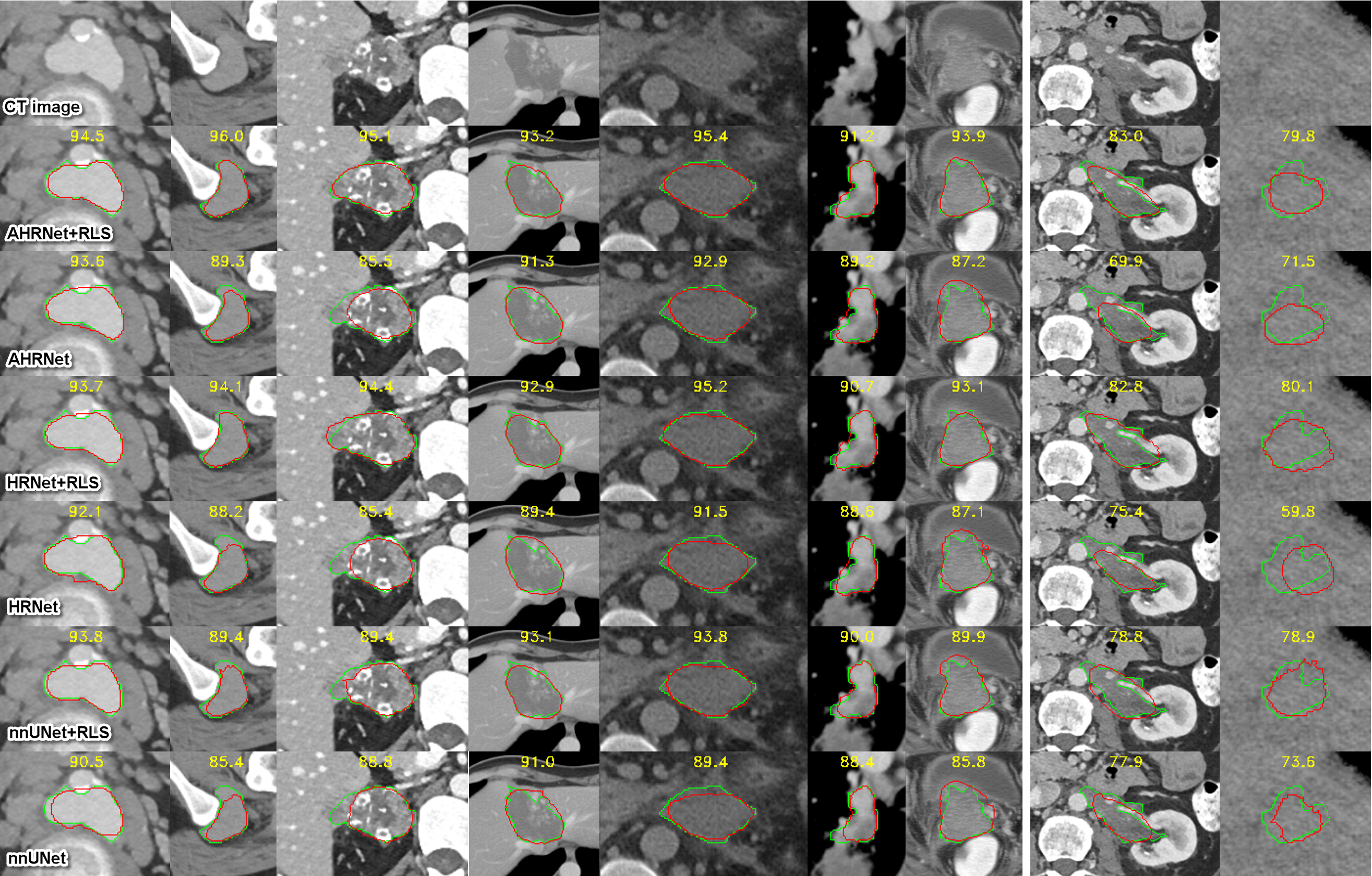} \\
	\end{minipage}
	\caption{Visual examples of results produced by different methods. The green and red curves represent the manual annotations and automatic segmentation results, respectively. The yellow digits indicate Dice scores. The last two columns give two failure cases.}
	\label{fig:result}
	\vspace{-5mm}
\end{figure}

Fig. \ref{fig:result} shows some {\bf visual examples} of lesion segmentation results by different methods for qualitative comparisons. As can be seen, 1) our lesion segmentation results are closer to the manual annotations than others, suggesting that our AHRNet model has desirable capability to learn more comprehensive features for distinguishing pixels nearby the lesion boundaries. 2) When using RLS, the results produced by all methods are closer to the manual annotations than the ones without RLS. Through optimizing the regional level set loss,  models can push or pull the segmentation results to improve the alignment with lesion boundaries. 3) When lesions have highly irregular shapes and blurry boundaries, all methods cannot segment them well, as shown in the last two columns of Fig. \ref{fig:result}. Beyond the weakly-supervised learning means, using a large number of fully- annotated data for training may alleviate these limitations.

\noindent \textbf{Ablation Studies.}
Table \ref{tab:ablation} lists quantitative comparisons of using different configurations to construct models for lesion segmentation. From Table \ref{tab:ablation}, when gradually introducing these components, \ie the decoder (DE), the dual-attention module (DA), and the scale attention module (SA), into the baseline (HRNet) sequentially, the performance is also improved accordingly. This indicates that these design options are useful to learn more representative features for lesion segmentation. When adding our RLS loss for training, it brings the largest performance gain (seeing row 4 versus row 6), \eg the Dice score is improved from 91.5\% to 92.6\%. The importance of RLS in our entire framework is validated. We also compute $\ell_{rls}$ using the entire input image rather than the constrained region during training, denoted as LS. From row 5 and 6, RLS achieves better performance than LS, implying that using the constrained regions for $\ell_{rls}$ computation is more reliable and beneficial for model optimization.

\begin{table}[t!]
    \centering
	\caption{Results of different configurations on the DeepLesion test set.}
	\label{tab:ablation}
	\scriptsize
    \begin{tabu} to 0.9\textwidth {| X[2.5] | X[c] | X[c] | X[c] |}
        \hline
        \multicolumn{1}{|c|}{Configurations} & Precision & Recall & Dice \\ \hline
        \hline
        (1) \quad Baseline (HRNet \cite{wang2020deep}) & \textbf{97.5$\pm$3.2}  &  84.9$\pm$8.6 & 90.5$\pm$5.3  \\ 
        (2) \qquad + DE & 95.5$\pm$5.8  &  87.7$\pm$9.1 & 91.0$\pm$5.9  \\
        (3) \qquad + DE + DA & 95.1$\pm$6.1  &  88.4$\pm$8.1 & 91.3$\pm$5.4  \\
        (4) \qquad + DE + DA + SA & 97.0$\pm$3.7  &  87.0$\pm$8.3 & 91.5$\pm$5.1  \\
        (5) \qquad + DE + DA + SA + LS & 96.2$\pm$4.3  &  89.4$\pm$7.7 &  92.2$\pm$4.6  \\
        (6) \qquad + DE + DA + SA + RLS & 95.8$\pm$4.5  &  \textbf{90.2$\pm$7.4} & \textbf{92.6$\pm$4.3}  \\
        \hline
    \end{tabu}
    \vspace{-5mm}  
\end{table}

\section{Conclusions}
In this paper, we present an attention enhanced high-resolution network (AHRNet) and a regional level set (RLS) loss for accurate weakly-supervised universal lesion segmentation. Instead of directly using the deep image features extracted by HRNet, AHRNet is able to learn more representative high-resolution features for lesion segmentation by integrating a decoder and attention mechanism. Assisted with the RLS loss, AHRNet model can further distinguish the pixels nearby the lesion boundaries more accurately. Extensive experimental results demonstrate that our proposed methods bring in better and more robust lesion segmentation results; specifically, RLS improves the performance significantly.

\bibliographystyle{splncs04}
\bibliography{egbib}
\end{document}